\documentclass[letterpaper, 10 pt, conference]{ieeeconf}  

\IEEEoverridecommandlockouts                              

\usepackage{graphics} 
\usepackage{epsfig} 
\usepackage{mathptmx} 
\usepackage{times} 
\usepackage{amsmath} 
\usepackage{amssymb}  
\usepackage{xcolor}
\usepackage{subcaption}
\usepackage{color}
\usepackage{hyperref}

\newcommand{\reviewChanged}[1]{#1}

\usepackage{fancyhdr} 
\usepackage{setspace}

\title{\LARGE \bf
Skeleton-Based Transformer for Classification of Errors and Better Feedback in \reviewChanged{Low Back Pain} Physical Rehabilitation Exercises
}

\author{Aleksa Marusic$^{1}$, Sao Mai Nguyen$^{1}$ and Adriana Tapus$^{1}$
\thanks{*This work was supported by ENSTA Paris}
\thanks{$^{1}$Autonomous Systems and Robotics Lab, Computer Science and System Engineering (U2IS), ENSTA Paris, Institut Polytechnique de Paris, 828 Blvd des Maréchaux, 91120 Palaiseau, France, {\tt\small name.surname@ensta-paris.fr}}%
}

\begin{document}

\maketitle
\thispagestyle{fancy}
\chead{
\vspace{-30pt}
\texttt{
\begin{spacing}{0.9}
\scriptsize{
Marusic, A., Nguyen, S. M., and Tapus, A. (2025). Skeleton-Based Transformer for Classification of Errors and Better Feedback in Low Back Pain Physical Rehabilitation Exercises.  International Conference on Rehabilitation Robotics. 
 }
 \end{spacing}
}}
 
\begin{abstract}
Physical rehabilitation exercises suggested by healthcare professionals can help recovery from various musculoskeletal disorders and prevent re-injury. However, patients’ engagement tends to decrease over time without direct supervision, which is why there is a need for an automated monitoring system. In recent years, there has been great progress in quality assessment of physical rehabilitation exercises. Most of them only provide a binary classification if the performance is correct or incorrect, and a few provide a continuous score. This information is not sufficient for patients to improve their performance. In this work, we propose an algorithm for error classification of rehabilitation exercises, thus making the first step toward more detailed feedback to patients. We focus on skeleton-based exercise assessment, which utilizes human pose estimation to evaluate motion. Inspired by recent algorithms for quality assessment during rehabilitation exercises, we propose a Transformer-based model for the described classification. Our model is inspired by the HyperFormer method for human action recognition, and adapted to our problem and dataset. The evaluation is done on the KERAAL dataset, as it is the only medical dataset with clear error labels for the exercises, and our model significantly surpasses state-of-the-art methods. Furthermore, we bridge the gap towards better feedback to the patients by presenting a way to calculate the importance of joints for each exercise.
\end{abstract}

\section{INTRODUCTION}

Physical rehabilitation plays a crucial role in helping patients recover from injuries, surgeries, and various medical conditions that affect their movement and functional abilities. 
Low back pain (LBP), which is the focus of our research, is a leading cause of disability globally, affecting over 50\% of the population at some point in their lives. Consequently, healthcare providers face substantial challenges in managing the increasing number of LBP patients \cite{Wu20}. 
The success of rehabilitation exercises depends significantly on how accurately and consistently they are performed. 
Throughout the weeks or even months of a rehabilitation program, patients often have to carry out exercises at home without direct oversight from healthcare professionals.
The absence of supervision and timely feedback from healthcare providers diminishes patient engagement in rehabilitation. Additionally, incorrect movements can not only delay the rehabilitation process but also increase the risk of further injuries \cite{Bassett2007HomeBasedPT}. 
Thus, there is a critical need for accurate and automated movement analysis methods to support patients and clinicians in monitoring rehabilitation exercises.




The primary objective of automated monitoring systems for physical rehabilitation is to identify the activity being performed, evaluate its quality, and offer detailed information on errors and potential improvements. Current movement analysis methods in rehabilitation typically provide a general quality score that reflects how well an exercise is executed \cite{Sardari2023, marusic2023analysis, mourchid2023dstgcn} 
While these scoring systems provide useful feedback, they represent only the initial step toward a fully automated and practical monitoring system.


A more detailed analysis, which not only evaluates the overall quality of movement but also identifies and localizes specific errors, would be highly beneficial for both patients and clinicians. This detailed feedback can help patients adjust their movements more precisely, keep them motivated throughout the rehabilitation sessions, and allow clinicians to more accurately monitor patients' progress during in-home rehabilitation \cite{Devanne2017ICHRH, Devanne2018IICRC}.


Human activity analysis is an evolving field that tackles the complex challenge of interpreting body movements based on data collected from various sources such as videos, images, or wearable sensors. 
\reviewChanged{While using data from wearable sensors directly as input is the dominant approach, recently, human pose estimation  to estimate joint coordinates from videos has become a promising approach to analyze human movement.}
Given that skeletons can be naturally modeled as graphs, it is not surprising that Graph Convolutional Networks (GCNs) have become a leading method for skeleton-based action recognition. By representing the body as a graph of joints and their spatial-temporal relationships, skeleton-based methods effectively capture movement dynamics and remain resilient to changes in appearance and environmental conditions, while capturing spatial and temporal relationships during exercise performance \cite{le2022harReview, shin2024harSurvey}. 

Furthermore, Transformer models, originally developed for natural language processing tasks, have shown remarkable success in various  applications, such as recognition of daily activities in smart homes \cite{Bouchabou2023smartHome}. 
Their ability to model long-range dependencies and capture contextual information makes them well-suited for analyzing sequential data, such as skeleton sequences in rehabilitation exercises \cite{xin2023transformer-review}.

For automated feedback allowing physical rehabilitation patients to improve their performance, this paper offers two key innovations: error classification and movement analysis. Unlike previous methods that focus solely on providing a quality score, our approach requires a more precise model, thus we utilize a skeleton-based transformer model. \reviewChanged{Through self-attention mechanism, our model is able to better learn spatial and temporal relations between skeleton joints.}
Specifically, our model is designed to:
\begin{itemize}
    \item Classify errors: 
    Identify and categorize different types of errors that may occur during the performance of rehabilitation exercises. 
    \item Biomechanical attention: 
     Identify the most important joints in human body skeleton for a specific movement, thus providing better feedback to the patient.
\end{itemize}

The remainder of this paper is organized as follows: 
Section \ref{sec:related} reviews related work on movement analysis in rehabilitation and skeleton-based action recognition models. After outlining the data we used in Section \ref{sec:data}, Section \ref{sec:algorithmy} describes the architecture of our skeleton-based transformer model and the methods for error classification and localization. Section \ref{sec:results} details our experimental setup and results, and discusses the implications of our findings. Finally, Section \ref{sec:conlcusion} summarizes our main contributions\reviewChanged{, acknowledges the limitations of our approach, and suggests potential directions for future research.}.

\section{RELATED WORK}
\label{sec:related}


\subsection{Skeleton-based Action Recognition}


Skeleton-based action recognition is a dynamic and rapidly evolving research area. Early studies depended on handcrafted features that utilized relative 3D rotations and translations between joints \cite{vemulapalli2014HumanAR,hussein2013}. However, the field has experienced significant advancements in recent years, largely driven by Deep Learning algorithms \cite{le2022harReview}. These Deep Learning methods for skeleton-based action recognition can be categorized into three primary groups, based on their approaches to extracting features from skeleton data for classification:

\begin{itemize}
    \item Recurrent Neural Networks (RNNs), which consider skeleton data primarily as a temporal sequence of continuous features. Some of the examples are \cite{du2015RNN} that introduced a hierarchical recurrent neural network, and \cite{Liu2017SkeletonBasedLSTM} that proposed Context-Aware Attention LSTM Networks.
    
    \item Convolutional Neural Networks (CNNs), \cite{Ke2017ANR,liu2017CNN} apply CNNs on pseudo-images obtained from skeleton data \cite{Ke2017ANR,liu2017CNN}. In this way, they capture spatial relations between joints in a frame.

    \item Graph Neural Networks (GNNs): Skeleton data naturally corresponds to a graph structure, with joints as vertices and bones as edges. As a result, Graph Convolutional Networks have become increasingly popular as they can extract both spatial relationships and be combined with temporal data \cite{yan2018STGCN, ye2020DynamicGCN}. 
\end{itemize}

\subsection{Physical Rehabilitation Assessment}


Qualitative exercise assessment is essential for effective home-based rehabilitation systems, aiming to provide patients with informative feedback and enhance their performance.
Early research on exercise evaluation applied traditional machine learning methods for classification, such as Adaboost, K-Nearest Neighbors (KNN), or Bayesian classifiers. Some approaches also utilized distance function-based models \cite{Ilktan2014, Houmanfar2016MovementAO}. Later studies adopted probabilistic models like Hidden Markov Models (HMMs) and Gaussian Mixture Models (GMMs) \cite{Devanne2017ICHRH, Vakanski2016}. While these models capture the stochastic nature of human motion and provide a quality score for movement accuracy, they do not fully exploit given information, such as joint or spatial connections between body parts.

Liao et al. \cite{Liao2020DLPRassessment} introduced a deep neural network model that generates quality scores for movements. They proposed a deep learning architecture for hierarchical spatio-temporal modeling, combining GMMs, CNNs, and LSTMs to compute a quality score. With the advent of Graph Neural Networks (GNNs), it is now possible to exploit spatial information through the skeletal graph structure. The authors in  \cite{deb2022graph} and \cite{Du2021AssessingPR} applied Graph Convolutional Networks (GCNs) to assess physical rehabilitation, achieving state-of-the-art results on popular datasets such as KIMORE and UI-PRMD. Additionally, Yu et al. \cite{Yu2022} employed an ensemble of two GCNs, one for position and one for orientation features of the skeletal joints, to further improve performance.

While existing methods primarily focus on quality scoring, a few studies have explored error classification and localization. For example, dynamic time warping (DTW) was used in \cite{chen2016real} to compare patient movements with reference movements and identify phases with significant deviations. Devanne et al. \cite{Devanne2018IICRC} added on top of DTW, a GMM model to analyse the likelihood per body part and per time segment to localize the error. However, its accuracy remains modest. On the other hand, our approach leverages the power of Deep Learning and Transformer models to capture the spatial-temporal dynamics of movements and provide actionable feedback with error classification and localization.
        
In the following sections, we describe our methodology in detail, including the architecture of our skeleton-based transformer model and the approaches for error classification and movement analysis.



\section{Data Processed}
\label{sec:data}

\reviewChanged{
\subsection{Dataset}

We evaluate our model on the Keraal dataset \cite{Nguyen2024IJCNN} collected during a clinical trial \cite{Blanchard2022BRI}, as it is the only one of the available rehabilitation datasets that fits into our problem setting and has error labels. TRSP \cite{trsp2017} is the only other rehabilitation dataset, to our knowledge, with error labels, but they only use two types of simple reaching motions. 

In the Keraal dataset, participants performed each of 3 predefined exercises, and the movements were recorded with Microsoft Kinect V2 and Vicon 
Recordings are annotated by physiotherapists in detail, with assessment of correctness, recognition of errors, and spatio-temporal localization of errors. In our model we used Kinect data as input, since it is a non-invasive system we would like to focus on.

The three exercises, chosen in agreement with medical experts, are namely torso rotation, flank stretch, and hiding face, which are illustrated in Figure \ref{fig:exercises}. For each exercise therapists identified three most common errors, and each labeled recording was classified either as correct or as one of these 3 errors. As can be seen on Figure \ref{fig:errors_descriptions}, the common mistakes for exercise (a) torso rotation are: 1) Arms are not raised enough, 2) The torso’s rotation is not sufficient, 3) The body is leaned on the side; for  exercise (b) flank stretch: 1) Opposite arm is not along the body, 2) Body is not tilted, 3) The above arm is not bent; and for   exercise (c) hiding face: 1) Arms are not raised enough, 2) Arms are not outspread enough, 3) Arms are not raised enough.

\begin{figure}
\centering
\includegraphics[width=0.8\linewidth]{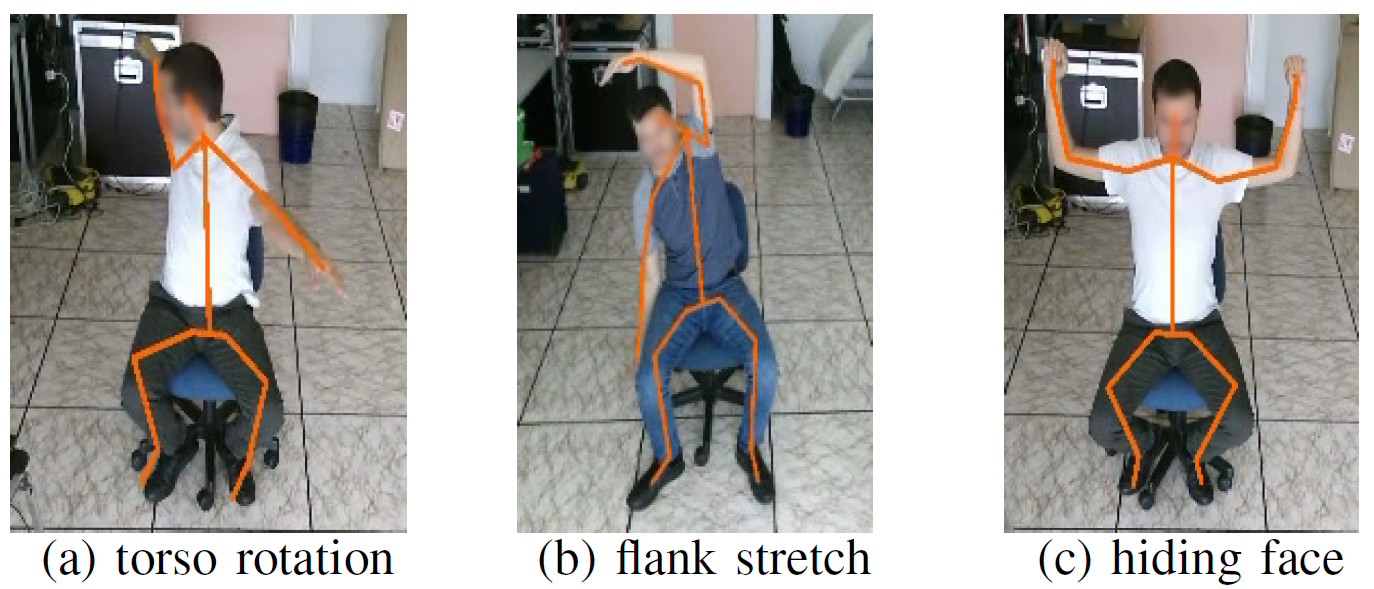}
\caption{The three rehabilitation exercises in the Keraal dataset. Image sourced from \cite{Nguyen2024IJCNN}.} 
\label{fig:exercises}
\vspace{-0.5em}
\end{figure}

\begin{figure}
\centering
\includegraphics[width=\linewidth]{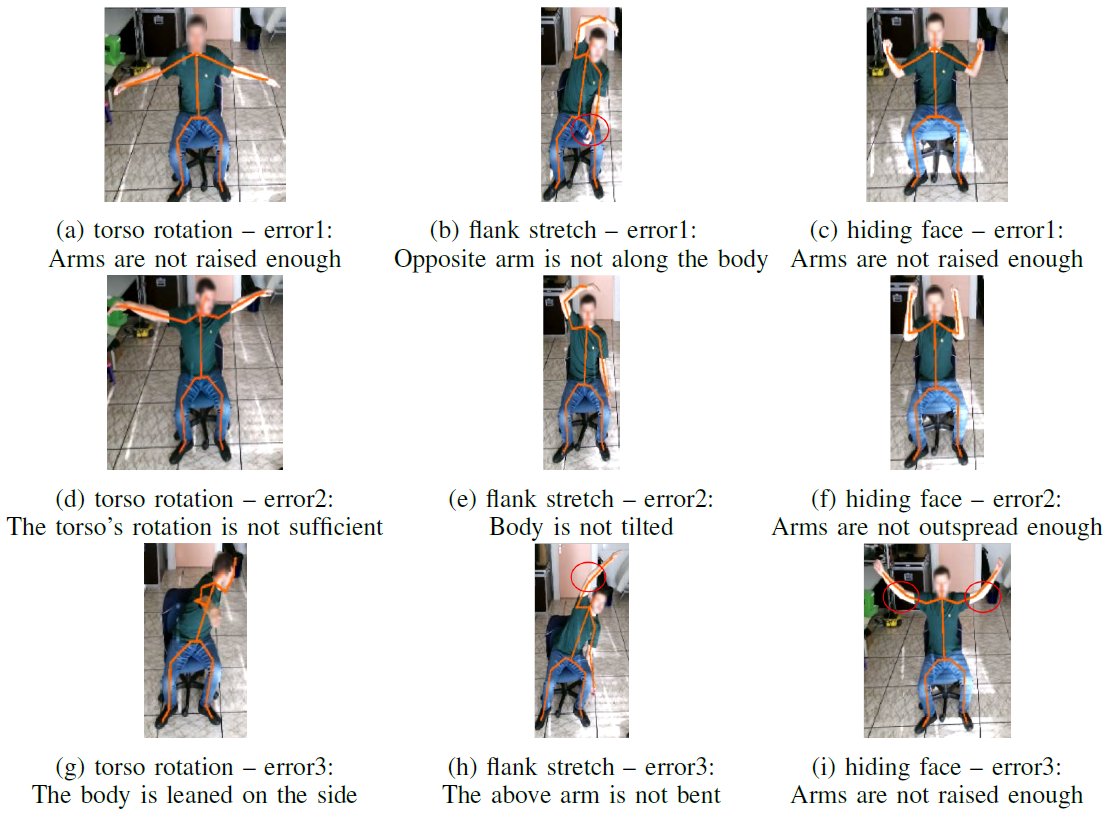}
\caption{Errors descriptions for all three exercises in the  Keraal dataset. Image sourced from \cite{Nguyen2024IJCNN}.} 
\label{fig:errors_descriptions}
\vspace{-1.5em}
\end{figure}

The Keraal dataset was recorded during a clinical rehabilitation study that included Low Back Pain patients, aged 18 to 70 years. The dataset includes recordings from both healthy subjects and 12 rehabilitation patients, from three groups of participants:
\begin{itemize}
    \item Group 1 where patients performed exercises
    \item Group 2 where healthy participants performed exercises without specific intructions
    \item Group 3 where trained healthy participants performed exercises while simulating errors.
\end{itemize}
Groups 1 and 2 were labeled by physiotherapists, while Group 3 was implicitly labeled as participants simulated errors.

\subsection{Data representation}

The data of the Keraal dataset recorded exercises at a fixed frame rate, capturing the spatial information of key skeletal joints in the human body, either in two or three dimensions. The data include joint positions and sometimes their orientations (in our case we use only 3D joint positions). 

Consider a tracking system that monitors \( V \) key points (joints) on a person's body. When someone performs one repetition of an exercise, the system records data across a certain number of time frames. Since that number can vary from recording to recording even for the same exercise type, we interpolate each recording to \( T \) timeframes. 
For each frame, we collect the positions of all \( V \) joints in a vector \( x(i), i \in [1 \ldots T] \). This vector has the dimension \( D = V \times C \), where \( C \) equals either 2 or 3, depending on whether we are working with 2D or 3D data. When we combine these vectors from all frames, we create a three-dimensional tensor \( X \in \mathbb{R}^{T \times V \times C} \), which represents the complete movement data from a single repetition.

\subsection{Evaluation scenarios}

Our goal is to classify a performed movement into one of four categories: correct, error1, error2, or error3. To achieve this, we train a separate model for each exercise. The primary metrics used for evaluation are the F1 score and accuracy. 

We consider three cases:
\begin{itemize}
    \item Scenario 1: Initial scenario where we train the models on 
    trained professionals simulating errors (group 3), and evaluate on healthy subjects and patients (groups 2A and 1A). This setup is critical as it demonstrates the model's performance on unseen data, closely simulating real-life application conditions.
    \item Scenario 2: In the second scenario, data from all three groups are combined and split into training and testing sets, in proportion 80:20. Also, it is important to note that the split is stratified, meaning the proportion of class labels is approximately the same in both training and test splits. This second scenario highlights the model's full potential by showing the level of performance achievable with sufficient data for both training and validation.
    \item Scenario 3: The scarcity of labels for certain exercises in the first scenario prompted us to design a third scenario. We still wanted to have unseen patient data in the test set to simulate real-life conditions, but we added a portion of healthy participants in the test set as well, so that we better evaluate our model. In this setup, the test set is made from patient data (group 1) and 15\% of the combined data from groups 2 and 3, while the training set consists of the remaining 85\% of the healthy participants' data (split is stratified in this case as well). 
\end{itemize}
}

\section{Algorithm}
\label{sec:algorithm}

\subsection{Self-attention}
\label{sec:self-attention}

Self-attention in Transformers was originally developed to identify and map relationships between words in a text, whether they are close together or far apart in a sentence. 
This mechanism helps the network prioritize important parts of the input data \cite{Bahdanau2014NeuralMT, vaswani2017attention}.

We can adapt this same concept to analyze human movement patterns captured through skeletal tracking. Just as Transformers process relationships between words in a text, we can process relationships between body joints in motion. Individual joint positions can be seen as words and each frame of movement as a sentence. By applying the self-attention mechanism, we can more effectively learn both local (between nearby joints or sequential frames) and distant relationships (between far-apart joints or temporally distant frames) in the motion sequence.

Consider a given input sequence \( X = (x_1, x_2, \dots, x_n) \) where each \( x_i \) represents an input token. Each token \( x_i \) is transformed into three vectors: Query vector \( q_i \), Key vector \( k_i \), and Value vector  \( v_i \). These vectors are obtained by applying learned projection matrices 
to each token, transforming them into new representations suited for computing attention. Essentially, these projections allow the model to learn more complex spatial relations of the input data. 

For each token \( x_i \), its attention score is computed relative to every other token in the sequence \( x_j \), including itself. The attention score is calculated by applying a softmax function on the dot product \( A_{ij} \) between the query vector \( q_i \) of token \( i \) and the key vector \( k_j \) of token \( j \): 

   \begin{equation}
   A_{ij} = q_i \cdot k_j^\top  
   \label{eq1}
   \end{equation}
   
To capture more complex relationships, an extension called Multi-Head Self-Attention (MHSA) is added \cite{vaswani2017attention}.
With multiple attention heads, the model learns different kinds of spatial relations between the joints,
in parallel.

\subsection{Model architecture}
\label{sec:model_architecture}

The main processing idea is inspired by a novel algorithm - Hyperformer \cite{hyperformer2023}. It has achieved state-of-the-art results on well-known human action recognition (HAR) benchmarks, proving it can learn complex relations between joints in the movement. However, unlike HAR datasets, we are not interested in distinguishing different types of actions, but different types of errors in one specific action. Thus, we are looking for much more subtle differences in motions during the same actions and trying to classify each motion accordingly. We train one model per exercise (action), compared to HAR algorithms which train one model in total to classify actions.

Furthermore, we have only a few classes (for errors), and even more importantly, very limited medical data to train the model, which emphasizes the need for a better understanding of spatio-temporal relations between the joints. That was another reason to choose Hyperformer as our basis, since it is one of the smallest HAR algorithms available (by number of trainable parameters).

\begin{figure}
\centering
\includegraphics[width=\linewidth]{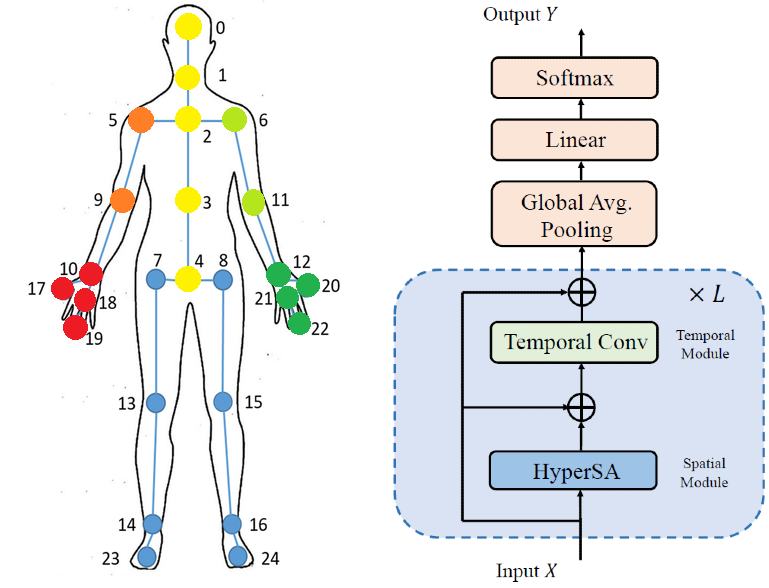}
\caption{Groups of skeletal joints (hypergraphs) on the left and model overview on the right. Right part taken from \cite{hyperformer2023} 
} 
\label{fig:hypergraph_layer}
\vspace{-1.5em}
\end{figure}

One of the key novelties of Hyperformer is utilizing hypergraphs - dividing the initial skeleton graph into subgraphs in order to obtain more precise relations between the joints. In our paper we propose a different split of the skeleton graph, more tailored to our problem setting. We split 25 joints into 6 different groups: left and right forearm and hand (with wrists and fingers), legs and spine with head, as can be seen on the left part of Figure \ref{fig:hypergraph_layer}. We give particular importance to arms as they are one of the key parts in all exercises from the dataset used, as opposed to legs which do not move much during the exercises. 

The overall model architecture can be seen on the right part of Figure ~\ref{fig:hypergraph_layer} 
The input data is passed through a certain number of layers (10 in our case). Each layer consists of a custom self-attention layer followed by Multi-Scale Temporal Convolution. Global pooling, a linear activation and softmax functions are added at the end. The model is trained via backpropagation through a cross-entropy loss function by comparing $\hat{\mathbf{y}}$ with the true label $\mathbf{y}$ - a class labeled by the physiotherapists.

\begin{figure}
\centering
\includegraphics[width=\linewidth]{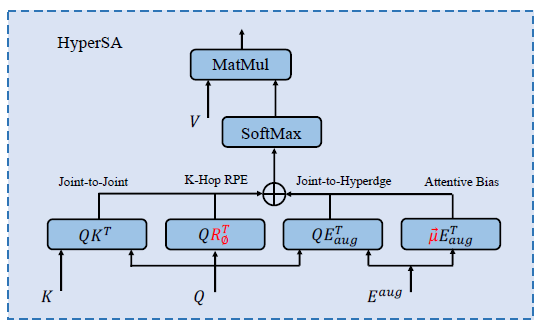}
\caption{Overview of the hyper self-attention module \cite{hyperformer2023}.} 
\label{fig:hyperSA}
\vspace{-1.0em}
\end{figure}

The main part of the layer is a custom self-attention module, as depicted in Figure \ref{fig:hyperSA}. It consists of 4 parts 
:
\begin{itemize}
    \item Joint-to-joint attention. Projection matrices used to get $q, k, v$ vectors are obtained with 2D convolutions. Attention is calculated as described in \ref{sec:self-attention}. 
    \item Joint-to-subgraph attention. Learns relations between joints and subgraphs, calculated the same way
    \item Relative positional embedding. Incorporates structural information of the skeleton, obtained with the Shortest Path Distance between the joints
    \item Attentive bias. Assigns the same amount of attention to each joint that belongs to a certain subgraph, enforcing independence of the query position.
\end{itemize}

\section{RESULTS \& DISCUSSION}
\label{sec:results}

In this section, we present the results of comprehensive comparative experiments conducted to assess the performance of our model. We provide a comparison between our proposed approach and the current state-of-the-art, as well as some guidelines for further improvement.

\reviewChanged{
\subsection{Implementation details}

Python 3.10 and the PyTorch 2.3 + CUDA 12.1 framework were used to develop the model. The system is equipped with an Intel Core i9-9900KF Silver 4215R CPU, 32GB of RAM, and an Nvidia RTX 2080 Ti 11GB. The model was trained with the Adam optimizer for 600 epochs, with a learning rate set to \(25 \times 10^{-4}\), and using a batch size of 10. Performance measures were recorded for each run and averaged to ensure accuracy. Our code is based on the Hyperformer code repository and is available at \url{https://github.com/aleksamarusic/hyperphysio/}
}

\subsection{Comparison with state-of-the-art}


\reviewChanged{Table \ref{table:acc} shows that out model reaches higher accuracies in Scenario 1, outperforming the models benchmarked in \cite{Nguyen2024IJCNN}. }

\begin{table}[h]
\caption{Accuracies 
in percentage }
\small
\label{table:acc}
\begin{center}
\begin{tabular}{| c | c | c | c | c |} 
 \hline
Exercise & Ours & LSTM best & LSTM mean & GMM \\ 
 \hline
Torso rotation & \textbf{73.17 } & 64.44  & 53.89  & 27.78  \\
 \hline
Flank stretch & \textbf{64.10 } & 43.04  & 31.64  & 25.32  \\ 
 \hline
Hiding face & \textbf{74.28 } & 56.19  & 49.1  & 33.33 \\ 
 \hline
\end{tabular}
\end{center}
\vspace{-1.0em}
\end{table}


Additionally, we present confusion matrices for the classification of rehabilitation movements, comparing three described scenarios 



\begin{figure*}
     \centering
     \begin{subfigure}[b]{0.3\linewidth}
         \centering
         \includegraphics[width=\textwidth]{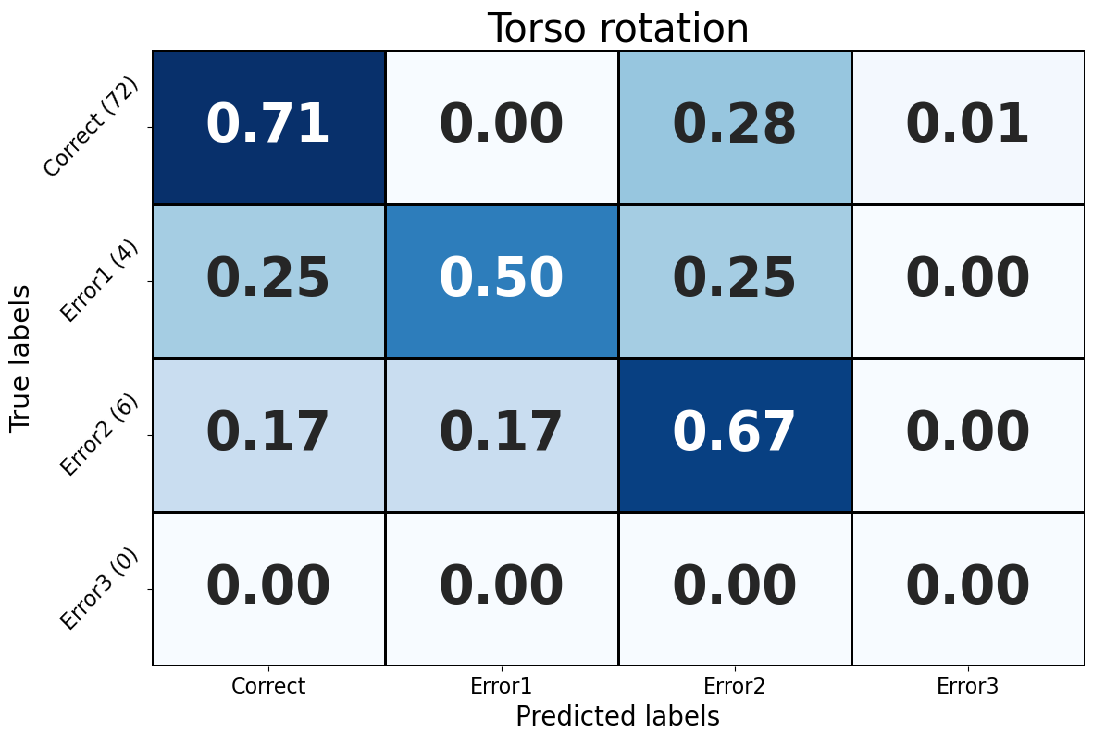}
         \caption{Torso rotation confusion matrix}
         \label{fig:conf_mat_ex1_first}
     \end{subfigure}
     \hfill
     \begin{subfigure}[b]{0.3\linewidth}
         \centering
         \includegraphics[width=\textwidth]{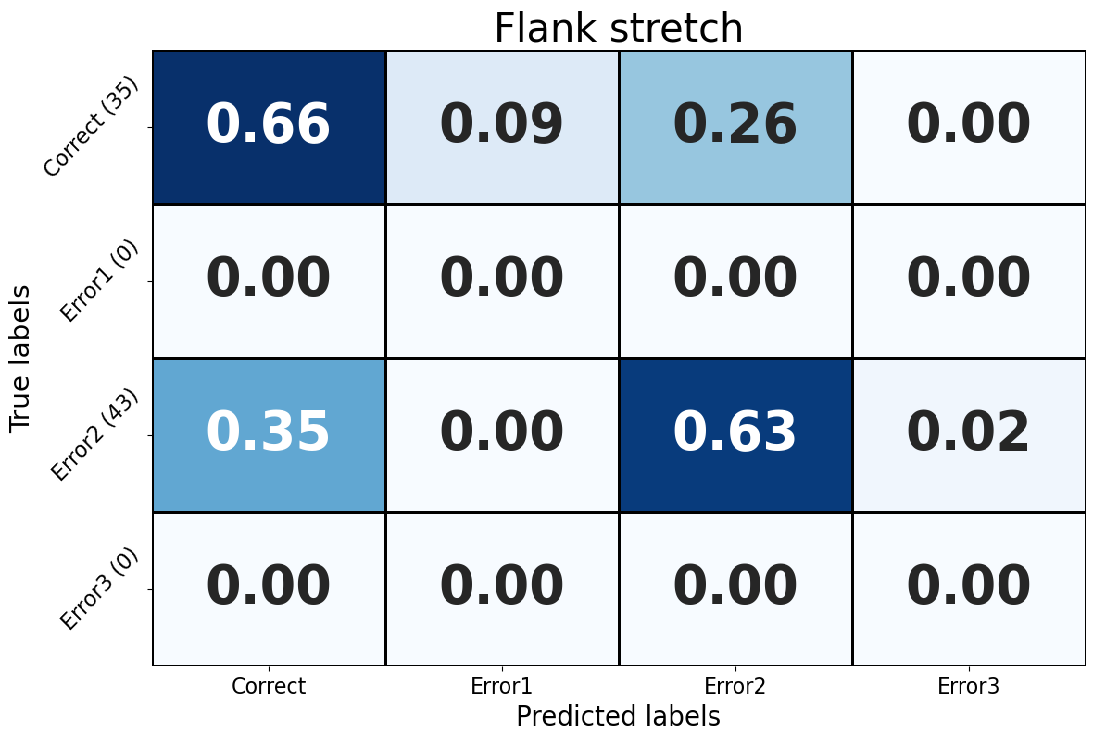}
         \caption{Flank stretch confusion matrix}
         \label{fig:conf_mat_ex2_first}
     \end{subfigure}
     \hfill
     \begin{subfigure}[b]{0.3\linewidth}
         \centering
         \includegraphics[width=\textwidth]{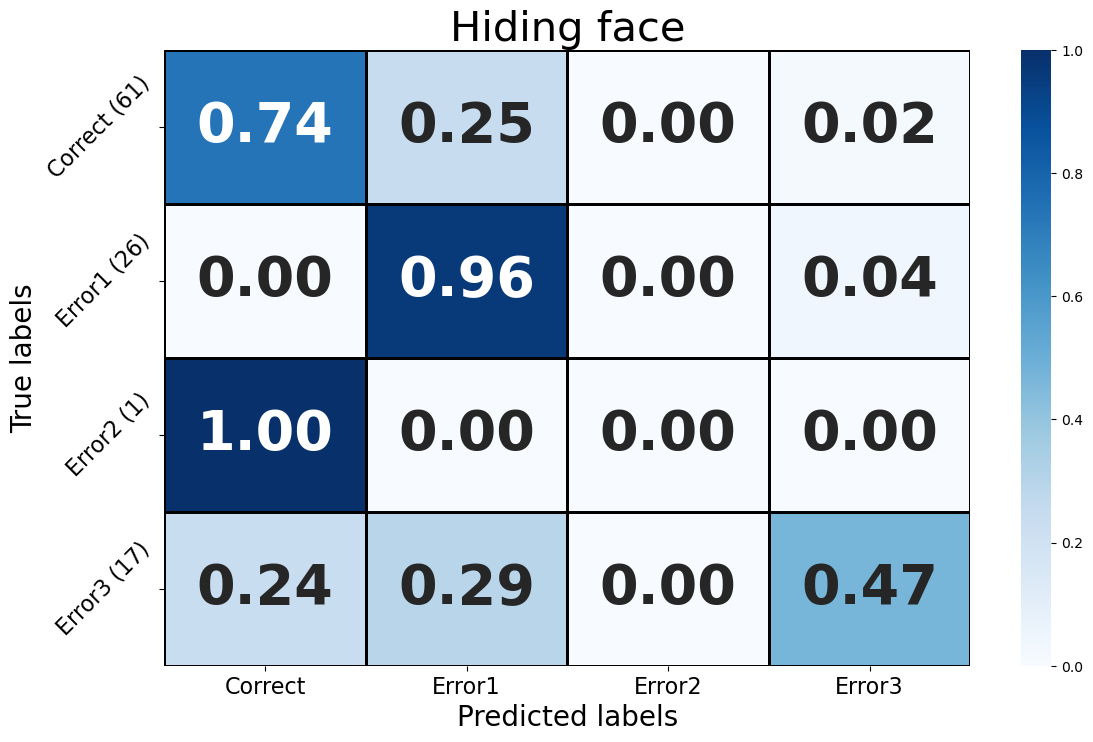}
         \caption{Hiding face confusion matrix}
         \label{fig:conf_mat_ex3_first}
     \end{subfigure}
  \vspace{-0.4em}
        \caption{Confusion matrices for each of 3 exercises for the first case (trained on group3 and tested on groups 2A and 1A
        ). Rows full of 0.00 mean that the corresponding error is not present in the test set.}
        \label{fig:conf_mat_all_first}
\end{figure*}

\reviewChanged{The confusion matrices for the first case are presented in Figure \ref{fig:conf_mat_all_first}, with true labels on the left and predicted labels on the bottom. The values in the matrices represent the percentage of exercises classified as a certain class. For instance, in Figure \ref{fig:conf_mat_ex1_first}, the first row indicates that 71\% of "Correct" exercises are classified as correct, 28\% as error2, and 1\% as error3. Consequently, the sum of any row in these matrices should be 1. Exceptionally, rows with all zeros mean that the corresponding error is not present in the test set. The value in brackets next to the left-side labels denote the frequency of the corresponding class in the test set.}

For the torso rotation exercise, it is very hard to evaluate the model performance as we have 72 "correct" labels, only 4 of error1 and 6 of error2, while error3 is not present in the test set.

Similarly, the scarcity of labels for errors 1 and 3 in the flank stretch exercise poses significant challenges for accurately evaluating the model's performance. Error2 exhibits some confusion with correct exercises, which can be explained by 
the general challenge to determine that the body is not tilted enough solely based on joint positions.

For hiding face, error2 has only 1 example in the test set, and we need more to properly evaluate. Also, it can be noticed that error3 shows misclassification with both correct and error1 categories. These can be also attributed to a general challenge as errors 1 and 3 are quite similar, and both are hard to distinguish from the correct performance as it is only a matter of a few pixels on the image whether they would be classified as one or the other.
These issues could potentially be mitigated with the inclusion of additional training and validation data, and with more precise data. 

\begin{figure*}
     \centering
     \begin{subfigure}[b]{0.3\linewidth}
         \centering
         \includegraphics[width=\textwidth]{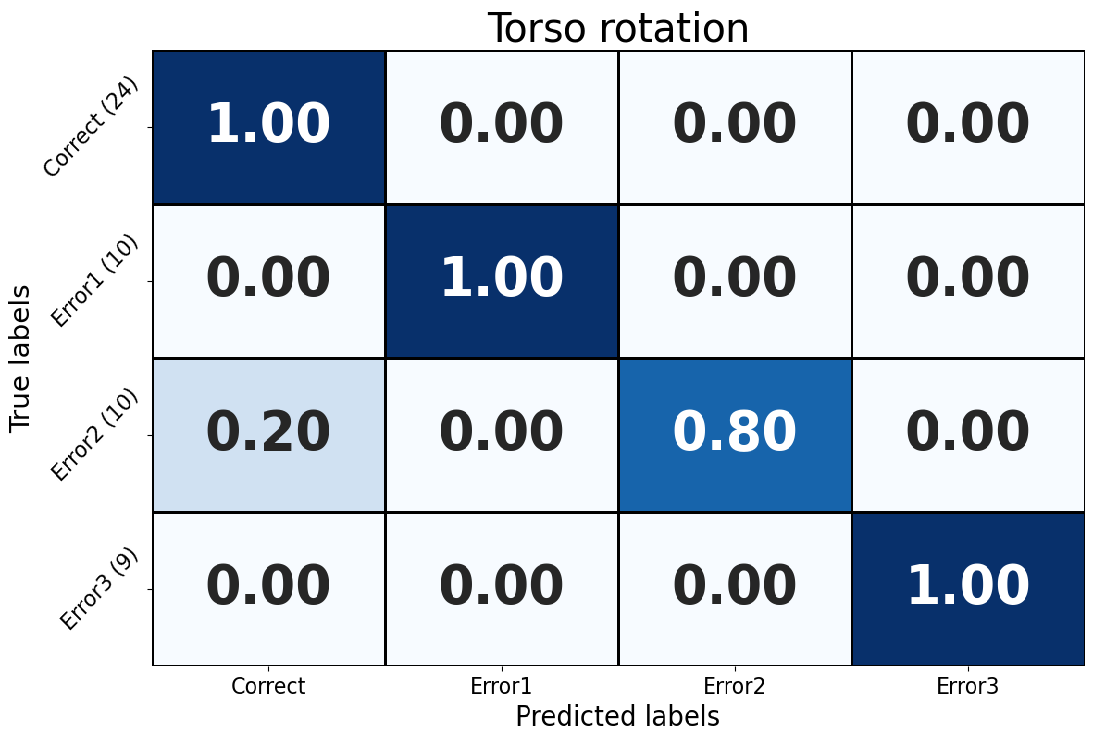}
         \caption{Torso rotation confusion matrix}
         \label{fig:conf_mat_ex1_second}
     \end{subfigure}
     \hfill
     \begin{subfigure}[b]{0.3\linewidth}
         \centering
         \includegraphics[width=\textwidth]{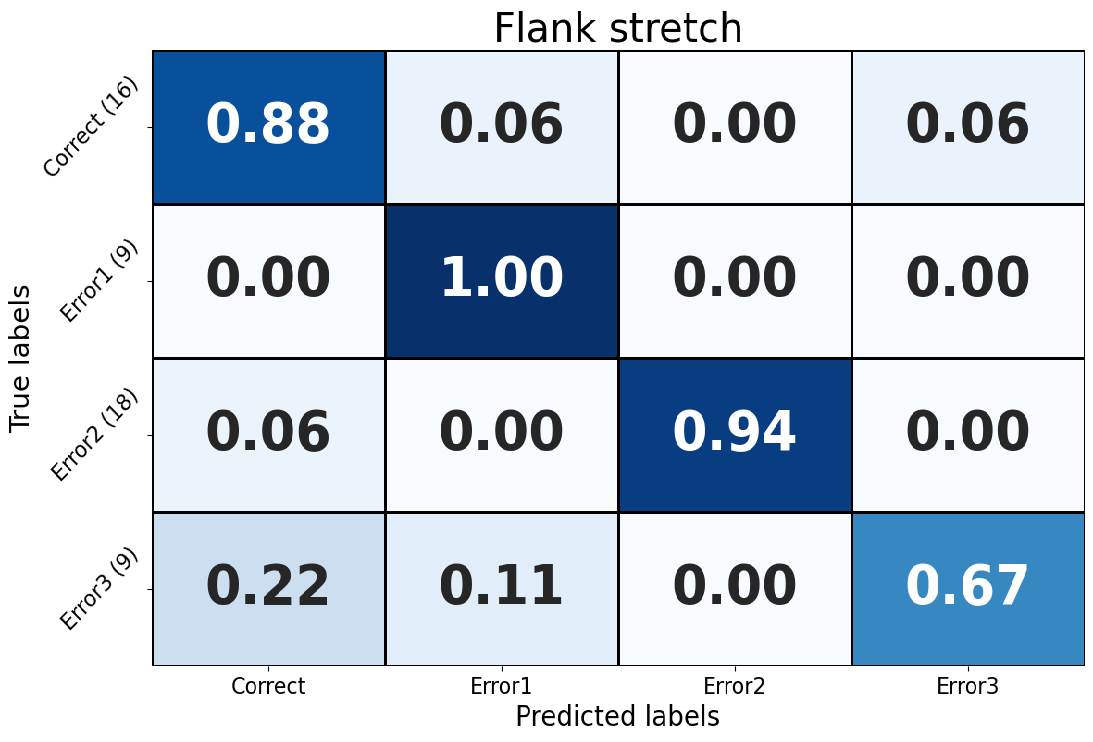}
         \caption{Flank stretch confusion matrix}
         \label{fig:conf_mat_ex2_second}
     \end{subfigure}
     \hfill
     \begin{subfigure}[b]{0.3\linewidth}
         \centering
         \includegraphics[width=\textwidth]{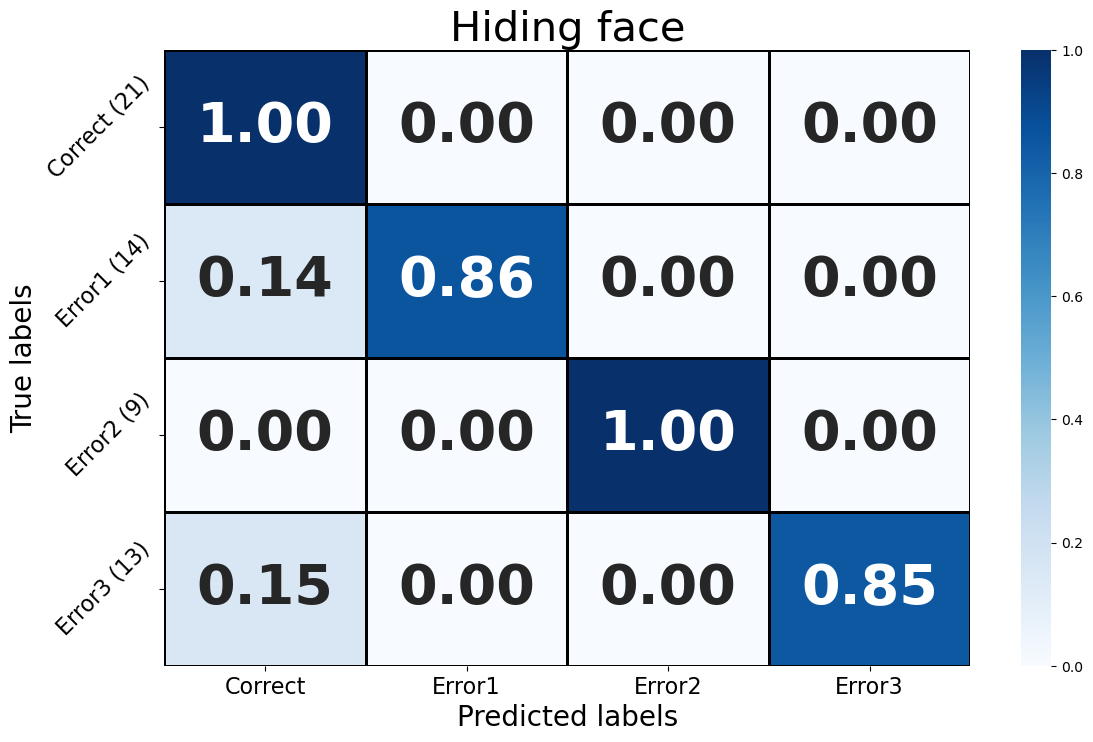}
         \caption{Hiding face confusion matrix}
         \label{fig:conf_mat_ex3_second}
     \end{subfigure}
  \vspace{-0.4em}
        \caption{Confusion matrices for each of 3 exercises for the second case (all groups combined and then data split into train and test sets).}
        \label{fig:conf_mat_all_second}
\end{figure*}

Confusion matrices for the second case can be seen in Figure \ref{fig:conf_mat_all_second}. As explained, we randomly sample 20\% of the dataset for testing, keeping the proportion of the labels in the test set. \reviewChanged{This scenario
obtained better performance, which confirms the
model's potential to perform significantly better when having more
training data.}

\begin{figure*}
     \centering
     \begin{subfigure}[b]{0.3\linewidth}
         \centering
         \includegraphics[width=\textwidth]{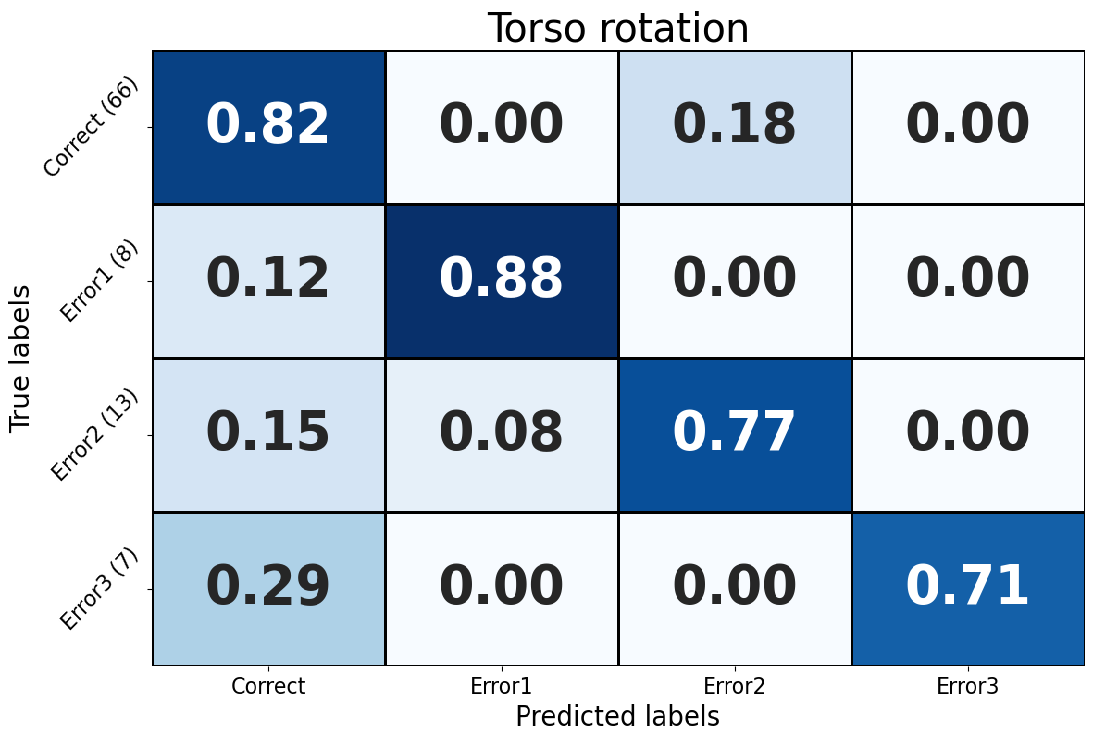}
         \caption{Torso rotation confusion matrix}
         \label{fig:conf_mat_ex1_third}
     \end{subfigure}
     \hfill
     \begin{subfigure}[b]{0.3\linewidth}
         \centering
         \includegraphics[width=\textwidth]{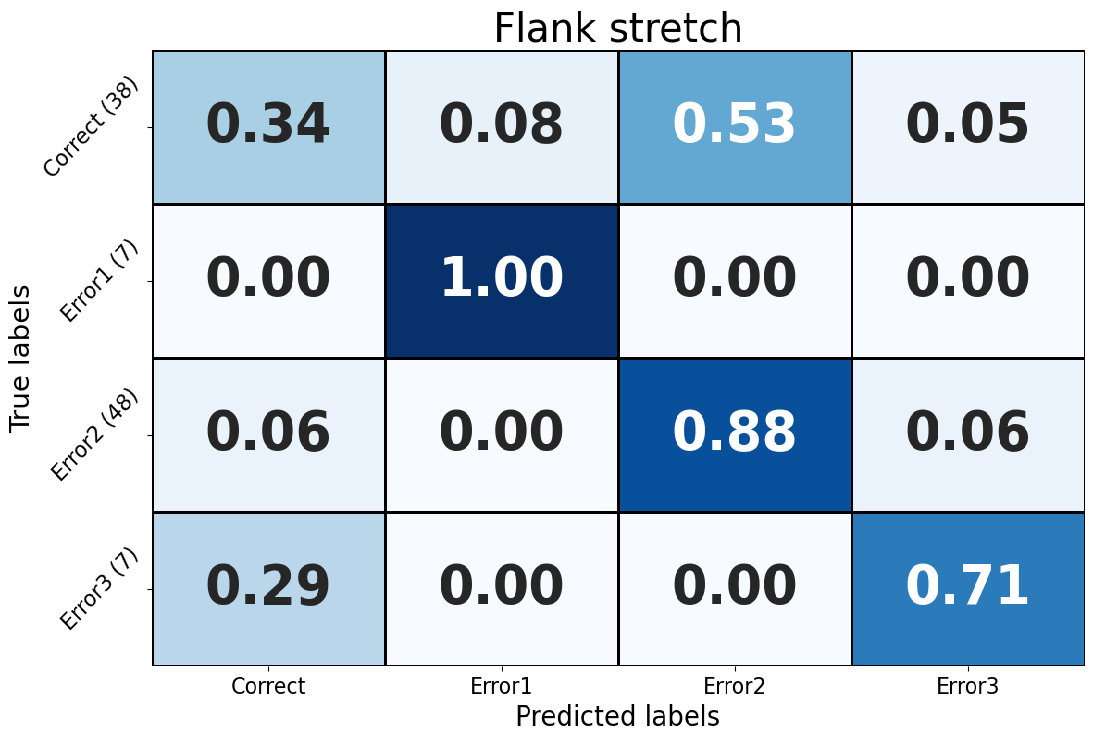}
         \caption{Flank stretch confusion matrix}
         \label{fig:conf_mat_ex2_third}
     \end{subfigure}
     \hfill
     \begin{subfigure}[b]{0.3\linewidth}
         \centering
         \includegraphics[width=\textwidth]{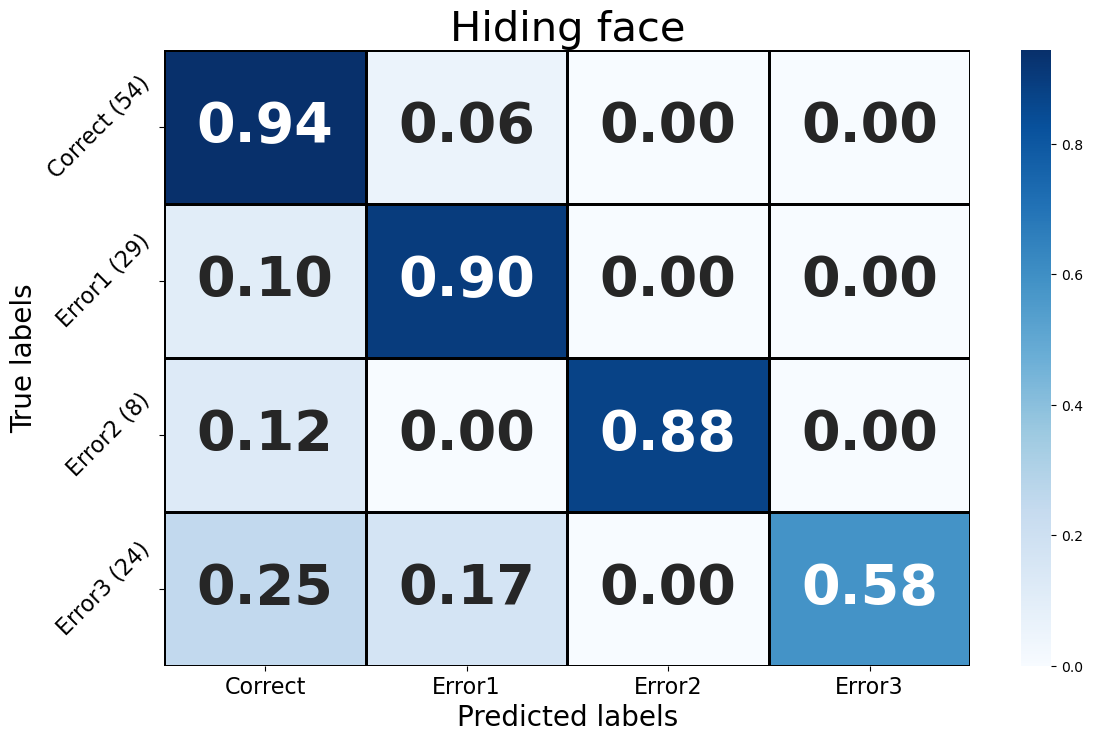}
         \caption{Hiding face confusion matrix}
         \label{fig:conf_mat_ex3_third}
     \end{subfigure}
    \caption{Confusion matrices on test for each of 3 exercises for the third case (trained on most \(85\%\) healthy (groups 2 \& 3) and tested on patients (group 1) and 15\% of groups 2 \& 3 combined.}
    \label{fig:conf_mat_all_third}
    \vspace{-1.5em}
\end{figure*}

The confusion matrices for the third case are shown in Figure \ref{fig:conf_mat_all_third}. While the majority of movements are correctly classified, some confusion can be observed between "correct" and "error3" in the flank stretch exercise 
. This may be attributed to the limited number of sequences labeled as correct by the medical expert and the subtle differences between correct movements and error2. The latter occurs when the body is not sufficiently tilted, a distinction that can be challenging to detect solely through skeleton joint data.

\subsection{Visualization of joints' importance}

The described visualization aims to analyze and compare the attention weights learned during the training and evaluation phase. By analyzing attention matrices, we can identify important joints in each exercise.

From our model, we can obtain the self-attention weights, which are quite comprehensive as we have weights for every layer, frame, and head (of multi-head attention) from every batch and input sequence. However, by averaging across these dimensions, we can obtain self-attention map that shows the attention weights for each body joint to every other body joint (since we have 25 joints, it is represented as \(25 * 25\) matrix). Further, the joint role can be computed by the column-wise summation over that map. Additionally, we also notice that is important to look at differences in self-attention maps obtained from correct and incorrect exercises. This way we can more easily obtain information on which joints play an important role in the exercise.

\begin{figure*}[ht]
\centering
\includegraphics[width=1.02\textwidth]{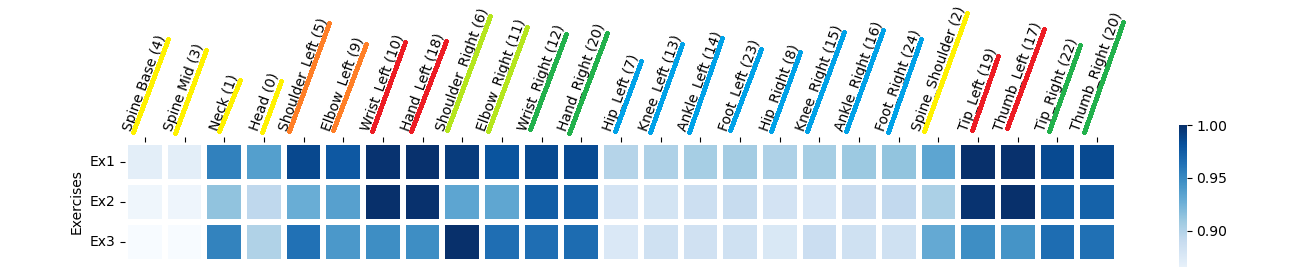}
\caption{An illustration of the joints' importance calculated from attention weights. Each row represents importance values for each of 25 skeleton joints during a specific exercise. Ex1 : Torso rotation, Ex2 : Flank stretch, and  Ex3 : Hiding face. 
Each column corresponds to a joint, designated by its name and its joint order (numbers in brackets next to labels), that are underlined with a color matching its group from Fig ~\ref{fig:hypergraph_layer}} 
\label{fig:joint_importance}
\vspace{-1.5em}
\end{figure*}

Looking at Figure \ref{fig:joint_importance}, 
we can notice that arm joints play an important role in all exercises, contrarily to the legs, as the exercises relate to the upper body. Additionally, Torso rotation and Hiding face give more importance to the angles of the shoulders and elbows, contrarily to Flank stretch. This is expected as the two former exercises require keeping the upper arm horizontal, while Flank stretch leaves the arm free, focusing more on the flank. 
These observations align with key joint patterns seen in the exercises (Figure \ref{fig:exercises}), confirming the ability of the model to provide accurate and useful feedback to patients.





\section{CONCLUSION}
\label{sec:conlcusion}

In this paper, we present a Transformer-based model for classifying errors in physical rehabilitation exercises. The model takes skeleton joint positions as input and classifies movement as either correct or one of the predefined errors. Inspired by the Hyperformer algorithm, our approach divides the skeleton graph into subgraphs to enhance the learning of complex spatial joint relations using the self-attention mechanism. 
\reviewChanged{We evaluate our model on the Keraal dataset, focusing on low back pain rehabilitation, which is the only dataset suitable for this error classification task. We can draw several conclusions:
\begin{itemize}
    \item Our model achieves significant improvements over benchmark algorithms and sets new state-of-the-art results. Our model better captures spatial and temporal relations with the self-attention module, and sets a new direction for automated rehabilitation assessment.
    \item We also calculate the importance of specific joints in performed exercises. This information can be used to detect which joints are wrongly moved, leading to enhanced feedback provided to the patients.
    \item Since we were interested in subtle differences in motions during the same exercise, it was necessary to train one model for each exercise type. 
    In a scenario in which patients are trained by a coach, the system knows which exercise is being performed and the focus was on the assessment method. A more comprehensive model integrating exercise detection, can be a future research direction. 
    \item One of the key limitations lies in the limited data available. While the Keraal dataset has detailed annotations, medical datasets are unfortunately limited in number.
    \item The imbalance of the dataset makes it harder to evaluate the model for underrepresented classes. Exploring data augmentation methods could lead to better classification and is a potential research direction for further studies.
    \item Some of the further research directions are temporal localization of errors, as well as more precise instructions on how to fix errors. These are essential for qualitative feedback to a patient.
    
\end{itemize}
}

\bibliographystyle{IEEEtran}
\bibliography{references}

\begin{thebibliography}{10}
\providecommand{\url}[1]{#1}
\csname url@samestyle\endcsname
\providecommand{\newblock}{\relax}
\providecommand{\bibinfo}[2]{#2}
\providecommand{\BIBentrySTDinterwordspacing}{\spaceskip=0pt\relax}
\providecommand{\BIBentryALTinterwordstretchfactor}{4}
\providecommand{\BIBentryALTinterwordspacing}{\spaceskip=\fontdimen2\font plus
\BIBentryALTinterwordstretchfactor\fontdimen3\font minus
  \fontdimen4\font\relax}
\providecommand{\BIBforeignlanguage}[2]{{%
\expandafter\ifx\csname l@#1\endcsname\relax
\typeout{** WARNING: IEEEtran.bst: No hyphenation pattern has been}%
\typeout{** loaded for the language `#1'. Using the pattern for}%
\typeout{** the default language instead.}%
\else
\language=\csname l@#1\endcsname
\fi
#2}}
\providecommand{\BIBdecl}{\relax}
\BIBdecl

\bibitem{Wu20}
A.~Wu, L.~March, X.~Zheng, J.~Huang, X.~Wang, J.~Zhao, F.~M. Blyth, E.~Smith,
  R.~Buchbinder, and D.~Hoy, ``Global low back pain prevalence and years lived
  with disability from 1990 to 2017: estimates from the global burden of
  disease study 2017,'' \emph{Annals of Translational Medicine}, 2020.

\bibitem{Bassett2007HomeBasedPT}
S.~F. Bassett and H.~Prapavessis, ``Home-based physical therapy intervention
  with adherence-enhancing strategies versus clinic-based management for
  patients with ankle sprains,'' \emph{Physical Therapy}, 2007.

\bibitem{Sardari2023}
S.~Sardari, S.~Sharifzadeh, A.~Daneshkhah, B.~Nakisa, S.~W. Loke, V.~Palade,
  and M.~J. Duncan, ``Artificial intelligence for skeleton-based physical
  rehabilitation action evaluation: A systematic review,'' \emph{Computers in
  Biology and Medicine}, 2023.

\bibitem{marusic2023analysis}
A.~Marusic, L.~Annabi, S.~M. Nguyen, and A.~Tapus, ``Analyzing data efficiency
  and performance of machine learning algorithms for assessing low back pain
  physical rehabilitation exercises,'' in \emph{2023 European Conference on
  Mobile Robots (ECMR)}, 2023.

\bibitem{mourchid2023dstgcn}
Y.~Mourchid and R.~Slama, ``D-stgcnt: A dense spatio-temporal graph conv-gru
  network based on transformer for assessment of patient physical
  rehabilitation,'' \emph{Computers in Biology and Medicine}, vol. 165, p.
  107420, 2023.

\bibitem{Devanne2017ICHRH}
M.~Devanne and S.~M. Nguyen, ``Multi-level motion analysis for physical
  exercises assessment in kinaesthetic rehabilitation,'' in \emph{2017 IEEE-RAS
  17th International Conference on Humanoid Robotics (Humanoids)}, 2017.

\bibitem{Devanne2018IICRC}
M.~Devanne, S.~M. Nguyen, O.~Remy-Neris, B.~Le~Gales-Garnett, G.~Kermarrec, and
  A.~Thepaut, ``A co-design approach for a rehabilitation robot coach for
  physical rehabilitation based on the error classification of motion errors,''
  in \emph{IEEE International Conference on Robotic Computing (IRC)}, Jan 2018,
  pp. 352--357.

\bibitem{le2022harReview}
V.-T. Le, K.~Tran-Trung, V.~T. Hoang, and H.~Chen, ``A comprehensive review of
  recent deep learning techniques for human activity recognition,''
  \emph{Intell. Neuroscience}, Jan. 2022.

\bibitem{shin2024harSurvey}
J.~Shin, N.~Hassan, A.~S.~M. Miah1, and S.~Nishimura, ``A comprehensive
  methodological survey of human activity recognition across divers data
  modalities,'' 2024.

\bibitem{Bouchabou2023smartHome}
D.~Bouchabou, J.~Grosset, S.~M. Nguyen, C.~Lohr, and X.~Puig, ``A smart home
  digital twin to support the recognition of activities of daily living,''
  \emph{Sensors}, 2023.

\bibitem{xin2023transformer-review}
W.~Xin, R.~Liu, Y.~Liu, Y.~Chen, W.~Yu, and Q.~Miao, ``Transformer for
  skeleton-based action recognition: A review of recent advances.''

\bibitem{vemulapalli2014HumanAR}
R.~Vemulapalli, F.~Arrate, and R.~Chellappa, ``Human action recognition by
  representing 3d skeletons as points in a lie group,'' \emph{2014 IEEE
  Conference on Computer Vision and Pattern Recognition}, pp. 588--595, 2014.

\bibitem{hussein2013}
M.~E. Hussein, M.~Torki, M.~A. Gowayyed, and M.~El-Saban, ``Human action
  recognition using a temporal hierarchy of covariance descriptors on 3d joint
  locations,'' ser. IJCAI 2013.

\bibitem{du2015RNN}
Y.~Du, W.~Wang, and L.~Wang, ``Hierarchical recurrent neural network for
  skeleton based action recognition,'' in \emph{2015 IEEE Conference on
  Computer Vision and Pattern Recognition (CVPR)}, 2015.

\bibitem{Liu2017SkeletonBasedLSTM}
J.~Liu, G.~Wang, L.~yu~Duan, K.~Abdiyeva, and A.~C. Kot, ``Skeleton-based human
  action recognition with global context-aware attention lstm networks,''
  \emph{IEEE Transactions on Image Processing}, vol.~27, pp. 1586--1599, 2017.

\bibitem{Ke2017ANR}
Q.~Ke, Bennamoun, S.~An, F.~Sohel, and F.~Boussa{\"i}d, ``A new representation
  of skeleton sequences for 3d action recognition,'' \emph{IEEE Conference on
  Computer Vision and Pattern Recognition}, 2017.

\bibitem{liu2017CNN}
M.~Liu, H.~Liu, and C.~Chen, ``Enhanced skeleton visualization for view
  invariant human action recognition,'' \emph{Pattern Recognition}, 2017.

\bibitem{yan2018STGCN}
S.~Yan, Y.~Xiong, and D.~Lin, ``Spatial temporal graph convolutional networks
  for skeleton-based action recognition,'' ser. AAAI'18/IAAI'18/EAAI'18.\hskip
  1em plus 0.5em minus 0.4em\relax AAAI Press, 2018.

\bibitem{ye2020DynamicGCN}
F.~Ye, S.~Pu, Q.~Zhong, C.~Li, D.~Xie, and H.~Tang, ``Dynamic gcn:
  Context-enriched topology learning for skeleton-based action recognition,''
  ser. MM.\hskip 1em plus 0.5em minus 0.4em\relax ACM, 2020.

\bibitem{Ilktan2014}
I.~Ar and Y.~S. Akgul, ``A computerized recognition system for the home-based
  physiotherapy exercises using an rgbd camera,'' \emph{IEEE Transactions on
  Neural Systems and Rehabilitation Engineering}, vol.~22, no.~6, pp.
  1160--1171, 2014.

\bibitem{Houmanfar2016MovementAO}
R.~Houmanfar, M.~Karg, and D.~Kuli{\'c}, ``Movement analysis of rehabilitation
  exercises: Distance metrics for measuring patient progress,'' \emph{IEEE
  Systems Journal}, vol.~10, 2016.

\bibitem{Vakanski2016}
A.~Vakanski, J.~Ferguson, and S.~Lee, ``Mathematical modeling and evaluation of
  human motions in physical therapy using mixture density neural networks,''
  \emph{Journal of Physiotherapy \& Physical Rehabilitation}.

\bibitem{Liao2020DLPRassessment}
Y.~Liao, A.~Vakanski, and M.~Xian, ``A deep learning framework for assessing
  physical rehabilitation exercises,'' \emph{IEEE Transactions on Neural
  Systems and Rehabilitation Engineering}, 2020.

\bibitem{deb2022graph}
S.~Deb, M.~F. Islam, S.~Rahman, and S.~Rahman, ``Graph convolutional networks
  for assessment of physical rehabilitation exercises,'' \emph{IEEE
  Transactions on Neural Systems and Rehabilitation Engineering}, vol.~30, pp.
  410--419, 2022.

\bibitem{Du2021AssessingPR}
C.~Du, S.~A. Graham, C.~A. Depp, and T.~Q. Nguyen, ``Assessing physical
  rehabilitation exercises using graph convolutional network with
  self-supervised regularization,'' \emph{International Conference of the IEEE
  Engineering in Medicine \& Biology Society}.

\bibitem{Yu2022}
B.~X. Yu, Y.~Liu, X.~Zhang, G.~Chen, and K.~C. Chan, ``Egcn: An ensemble-based
  learning framework for exploring effective skeleton-based rehabilitation
  exercise assessment,'' in \emph{Proceedings of International Joint Conference
  on Artificial Intelligence 2022}.

\bibitem{chen2016real}
X.~Chen, Y.~Liu, and Q.~Huang, ``Real-time error detection and feedback for
  physical rehabilitation exercises using kinect and dtw,'' 2016.

\bibitem{Nguyen2024IJCNN}
S.~M. Nguyen, M.~Devanne, O.~Remy-Neris, M.~Lempereur, and A.~Thepaut, ``A
  medical low-back pain physical rehabilitation dataset for human body movement
  analysis,'' in \emph{International Joint Conference on Neural Networks},
  2024.

\bibitem{Blanchard2022BRI}
A.~Blanchard, S.~M. Nguyen, M.~Devanne, M.~Simonnet, M.~L. Goff-Pronost, and
  O.~R{\'e}my-N{\'e}ris, ``Technical feasibility of supervision of stretching
  exercises by a humanoid robot coach for chronic low back pain: The r-cool
  randomized trial,'' \emph{BioMed Research International}, vol. 2022, pp.
  1--10, mar 2022.

\bibitem{trsp2017}
E.~Dolatabadi, Y.~X. Zhi, B.~Ye, M.~Coahran, G.~Lupinacci, A.~Mihailidis,
  R.~Wang, and B.~Taati, ``The toronto rehab stroke pose dataset to detect
  compensation during stroke rehabilitation therapy,'' 2017.

\bibitem{Bahdanau2014NeuralMT}
\BIBentryALTinterwordspacing
D.~Bahdanau, K.~Cho, and Y.~Bengio, ``Neural machine translation by jointly
  learning to align and translate,'' \emph{CoRR}, 2014. [Online]. Available:
  \url{https://api.semanticscholar.org/CorpusID:11212020}
\BIBentrySTDinterwordspacing

\bibitem{vaswani2017attention}
A.~Vaswani, N.~Shazeer, N.~Parmar, J.~Uszkoreit, L.~Jones, A.~N. Gomez,
  L.~Kaiser, and I.~Polosukhin, ``Attention is all you need,'' in
  \emph{Proceedings of the 31st International Conference on Neural Information
  Processing Systems}, 2017.

\bibitem{hyperformer2023}
Z.~Hu, V.~Guti\'{e}rrez-Basulto, Z.~Xiang, R.~Li, and J.~Z. Pan, ``Hyperformer:
  Enhancing entity and relation interaction for hyper-relational knowledge
  graph completion,'' 2023.

\end{thebibliography}
\end{document}